\documentclass[superscriptaddress, twocolumn, prl]{revtex4}

\usepackage{amsmath}
\usepackage{amssymb}
\usepackage{dsfont}
\usepackage{color}
\usepackage{hyperref}
\usepackage{graphicx}
\usepackage{epstopdf}
\usepackage{soul}
\usepackage{color}
\usepackage{braket}
\usepackage[normalem]{ulem}

\usepackage{xcolor}
\hypersetup{
  colorlinks   = true, %Colours links instead of ugly boxes
  urlcolor     = blue, %Colour for external hyperlinks%%% Email
  linkcolor    = blue, %Colour of internal links%%%% Fig.
  citecolor   = magenta %Colour of citations %%% Ref.
}

\begin{document}
\title{Observation of pair tunneling and coherent destruction of tunneling\\ in arrays of optical waveguides}

\author{Sebabrata~Mukherjee}
\email{snm32@hw.ac.uk}
\affiliation{Institute of Photonics and Quantum Sciences, School of Engineering $\&$ Physical Sciences, Heriot-Watt University, Edinburgh, EH14 4AS, United Kingdom}
\author{Manuel~Valiente}
\affiliation{Institute of Photonics and Quantum Sciences, School of Engineering $\&$ Physical Sciences, Heriot-Watt University, Edinburgh, EH14 4AS, United Kingdom}
\author{Nathan~Goldman}
\affiliation{Center for Nonlinear Phenomena and Complex Systems, Universit\'e Libre de Bruxelles, CP 231, Campus Plaine, B-1050 Brussels, Belgium} 
\author{Alexander~Spracklen}
\affiliation{Institute of Photonics and Quantum Sciences, School of Engineering $\&$ Physical Sciences, Heriot-Watt University, Edinburgh, EH14 4AS, United Kingdom}
\author{Erika~Andersson}
\affiliation{Institute of Photonics and Quantum Sciences, School of Engineering $\&$ Physical Sciences, Heriot-Watt University, Edinburgh, EH14 4AS, United Kingdom}
\author{Patrik \"Ohberg}
\affiliation{Institute of Photonics and Quantum Sciences, School of Engineering $\&$ Physical Sciences, Heriot-Watt University, Edinburgh, EH14 4AS, United Kingdom}
\author{Robert R. Thomson}
\affiliation{Institute of Photonics and Quantum Sciences, School of Engineering $\&$ Physical Sciences, Heriot-Watt University, Edinburgh, EH14 4AS, United Kingdom}

\begin{abstract}
We report on the experimental realization of a photonic system that simulates the one-dimensional two-particle Hubbard model. This analogy is realized by means of two-dimensional arrays of coupled optical waveguides, fabricated using femtosecond laser inscription. By tuning the analogous ``interaction strength", we reach the strongly-interacting regime of the Hubbard Hamiltonian, and demonstrate the suppression of standard tunneling for individual ``particles". In this regime, the formation of bound states is identified through the direct observation of pair tunneling. %This effect is emphasized through the suppression of standard tunneling for individual ``particles". 
We then demonstrate the coherent destruction of tunneling (CDT) for the paired particles in the presence of an engineered oscillating force of high frequency. The precise control over the analogous ``interaction strength" and driving force offered by our experimental system opens an exciting route towards quantum simulation of few-body physics in photonics.
\end{abstract}
%71.10.Fd	Lattice fermion models (Hubbard model, etc.)
%03.65.Ge	Solutions of wave equations: bound states
%42.82.Et	Waveguides, couplers, and arrays
%63.20.Pw	Localized modes
%03.65.Xp	Tunneling, traversal time, quantum Zeno dynamics
%32.80.Qk	Coherent control of atomic interactions with photons
%\pacs{71.10.Fd, 03.65.Ge, 42.82.Et, 63.20.Pw}
\maketitle
%%%%%%%%%%%%%%%   Introduction   %%%%%%%%%%%%%%%%
{\it Introduction.}  
Elucidating the physics of interacting electrons in crystalline solids constitutes one of the most challenging problems in modern physics, with direct implications for our understanding of quantum magnetism and superconductivity. Theoretical toy models, such as the Hubbard model~\cite{Hubbard1963electron}, can be used to examine these problems, but challenging issues can arise, especially in the intermediate to strong coupling regimes where perturbation theory fails. However, the one-dimensional Hubbard model is exactly solvable by means of Bethe Ansatz techniques~\cite{liebwu1968, caffarel1998hubbard}. In particular, the two-particle solution in the singlet sector -- the triplet sector is non-interacting -- shows both scattering and bound states for any value of the interaction strength. Remarkably, bound state solutions, known as doublons, exist even in the presence of repulsive interactions. This phenomenon can be understood in terms of the band gap, or the boundedness of the spectrum in the Hubbard model, which implies that two repulsively interacting particles, initially occupying the same lattice well, will have no available scattering energies to dissociate into. This can be further explained by an exact symmetry between attractive and repulsive interactions reported in~\cite{mosseri2000some, valiente2010, schneiderrosch2012}. These repulsively bound states were experimentally observed using both bosonic~\cite{winkler2006repulsively} and fermionic~\cite{strohmaier2010observation} particles in optical lattices. These experiments triggered an intense activity exploring the physics of few particles in optical lattices, including the two-body~\cite{petrosyan2007quantum, valientepetrosyan2008, valientepetrosyan2009, creffield2010coherent, javanainen2010dimer, zhang2012bound, longhi2013klein, qin2014statistics, bello2015long, preiss2015strongly, bello2016sublattice} and three-body~\cite{orsoburovskijolicoeur2009, valientepetrosyansaenz2010} problems, in which certain phenomena that have no analogy in free space occur. For example, the Mattis-Gallinar effect~\cite{mattisgallinar1984}, due to the absence of Galilean invariance, states that the effective mass of a lattice bound pair is higher than its free-space analogue, and was observed for excitons~\cite{cafolla1985}. Three-body composites in bosonic~\cite{valientepetrosyansaenz2010} or mass-imbalanced one-dimensional fermionic systems~\cite{orsoburovskijolicoeur2009} can also exist due to an effective exchange mechanism between a bound pair and a neighboring third particle. It is of great interest to identify these exotic few-body properties in the absence of a many-body environment. This is difficult using solid-state systems and challenging to probe using traditional cold-atom experiments. Recently, few-body physics was explored with cold atoms in the absence of an optical lattice~\cite{jochim2011, jochim2012};
%\cite{jochim2011, jochim2012, jochim2013, jochim2013-2, jochim2015}; 
using this approach,  a one-dimensional Heisenberg spin chain~\cite{volosniev2014, deuretzbacher2014} was realized using few atoms~\cite{jochim2015-2}.

Interestingly, the physics of two interacting quantum particles moving in a one-dimensional lattice can be simulated using a non-interacting system, such as photonic crystals operating in the linear optical regime~\cite{longhi2011optical, krimer2011realization, longhi2011photonic, Longhi2012coherent, corrielli2013fractional, rai2015photonic}.
This approach is based on a mapping, which we now briefly summarize; see also~\cite{noba2003dynamic, longhi2011tunneling}. Let us consider the standard Hubbard model (with $\hbar\!=\!1$) in one dimension,
\begin{equation}
\hat H\!=\!-J \sum_{s} \big(\hat a^{\dagger}_s \hat a_{s+1}+\hat a_{s+1}^\dagger \hat a_s \big)+\frac{U_0}{2} \hat n_s (\hat n_s-1), \label{Hubbard} 
\end{equation}
where $\hat a_{s}^\dagger$ ($\hat a_{s}$) is the creation (annihilation) operator for a particle at the $s$-th lattice site, $\hat n_s\!=\!\hat a_{s}^\dagger \hat a_{s}$ is the number operator, $J$ is the nearest-neighbor hopping amplitude and $U_0$ is the on-site interaction energy (note that $U_0\!>\!0$ for repulsive interactions). Considering only two particles in the system, one expands the two-body wave function in the Fock basis as $\ket{\psi(t)}\!=\!(1/\sqrt 2)\sum_{l, m} C_{l, m}(t) \ket{l, m}$, where $C_{l, m}$ is the probability amplitude for finding one particle at site $l$ and the other one at site $m$. The time-evolution of these coefficients is then determined by solving the equation
\begin{align}
i\frac{dC_{l, m}(t)}{dt}\!=\!&-J \big(C_{l+1, m}+ C_{l-1, m}+ C_{l, m+1}+ C_{l, m-1}\big) \nonumber \\
&+U_0 C_{l, m} \delta_{l, m},   \label{eq:single}
\end{align}
which results from the Schr\"odinger equation $i \partial_{t} \ket{\psi}\!=\!\hat H \ket{\psi}$. Importantly, Eq.~\eqref{eq:single} can be formally mapped into a Schr\"odinger equation that describes the dynamics of a \emph{single particle} in a two-dimensional lattice, with sites located at position $\boldsymbol{r}\!=\!a_x l \boldsymbol{1}_x + a_y m \boldsymbol{1}_y$, where $a_{x,y}$ denote the lattice
constants, and ($l$, $m$) are integers labeling the lattice sites in the 2D plane. In this picture, $J$ denotes the hopping amplitude along both spatial directions, and $U_0$ corresponds to an on-site potential, which only affects the sites located along the diagonal $l\!=\!m$. As a result, any non-interacting two-dimensional system described by Eq.~\eqref{eq:single} can be exploited to capture the physics of the two-body Hubbard problem [Eq.~\eqref{Hubbard}].

\begin{figure}[]
\includegraphics[width=8.6 cm]{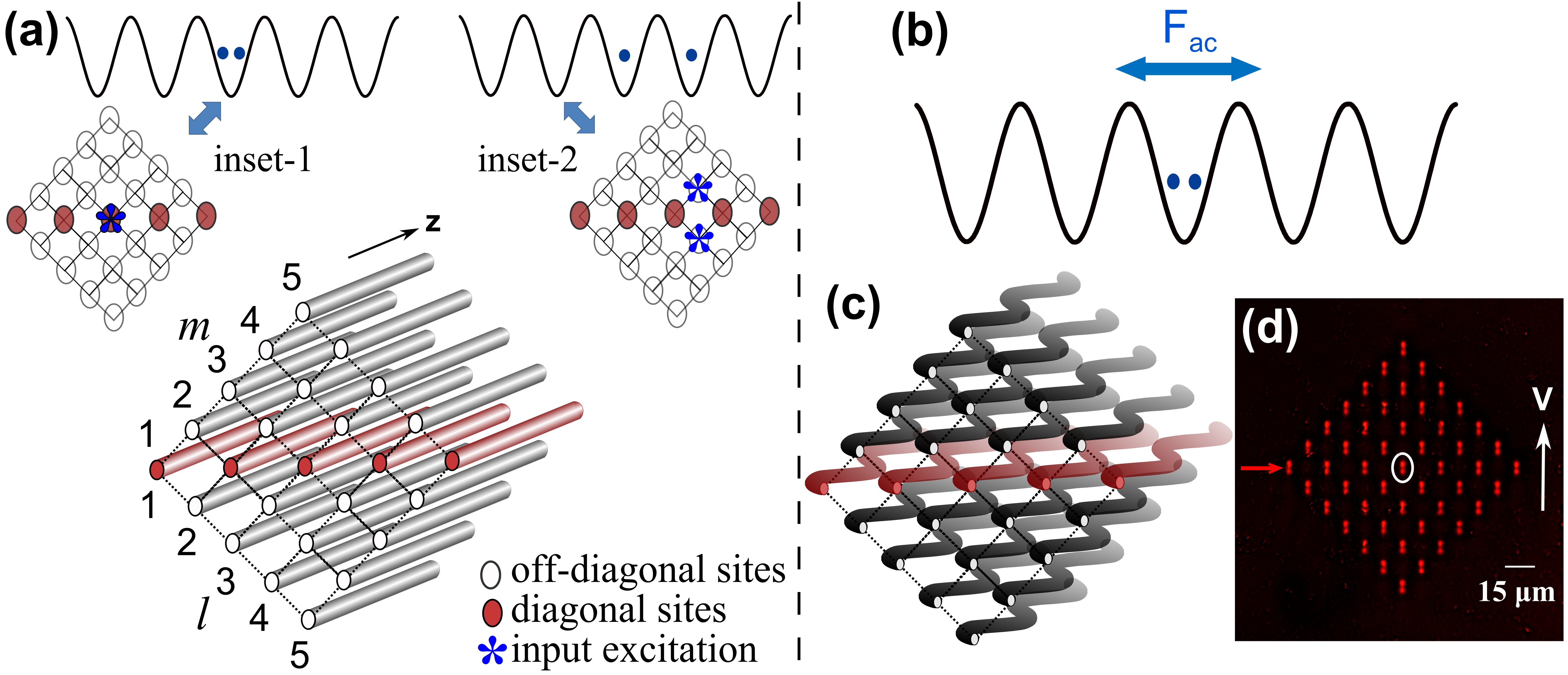}
\caption{(a) Schematic diagram of a photonic square lattice where the diagonal sites $l$=$m$ (shown in red) have a shifted propagation constant, $\Delta \beta$ [$\equiv\!U_0$, the on-site interaction energy] 
compared to the off-diagonal waveguides. The excitation of a diagonal waveguide at the input is the optical analog of two interacting particles at the same potential well of a one-dimensional lattice at time $t\!=\!0$ (inset-1). The excitation of the two off-diagonal waveguides corresponds to the particles located at different lattice sites at $t\!=\!0$ (inset-2); see also Ref.~\cite{krimer2011realization}.
(b) Schematic diagram of two interacting particles in a one-dimensional lattice driven by a sinusoidal force ($F_{ac}$). %with an external alternating force ($F_{ac}$); $U_0$ is the on-site interaction.
(c) A modulated photonic square lattice which mimics the dynamics of (b). 
(d) Micrograph of the facet of a laser-fabricated photonic square lattice. The lattice axes were rotated by 45$^{\circ}$ with respect to the vertical direction (V) to achieve $\kappa_x\!\approx\!\kappa_y\!=\!\kappa$~(see APPENDIX). %\cite{suppmat}. 
The circled waveguide was excited at the input for all measurements. The red arrow indicates the main diagonal ($l\!=\!m$).
}
\label{fig1}
\end{figure}

In this Letter, we explore this mapping by considering a photonic square lattice ($a_x, a_y\!=\!a$), a square array of coupled optical waveguides [Fig.~\ref{fig1}~(a)]. Recently, such photonic systems have been developed to study single-particle effects such as CDT in a double well~\cite{longhi2005coherent, della2007visualization}, Bloch oscillations~\cite{chiodo2006imaging}, dynamic localization~\cite{longhi2006observation, szameit2009polychromatic}, Landau-Zener tunneling~\cite{dreisow2009direct},  Anderson localization~\cite{schwartz2007transport}, chiral edge modes~\cite{Rechtsman:2013}, and the localized states associated with flat-band lattices~\cite{mukherjee2015observation, vicencio2015Observation, Mukherjee2015Rhombic}. Specific effects related to interacting systems have also been studied based on photonic lattices, see Refs.~\cite{longhi2011optical, krimer2011realization, longhi2011photonic, Longhi2012coherent, corrielli2013fractional, rai2015photonic}, suggesting an interesting alternative to cold-atom experiments~\cite{folling2007direct, preiss2015strongly}.

%%%%%%     The implementation and the mapping       %%%%%%%%
{\it Implementation and mapping.} 
We consider a straight photonic square lattice where the waveguides located along the main diagonal ($l\!=\!m$) have a shifted propagation constant ($\Delta \beta\!=\!\beta_{\text{off-diagonal}}\!-\!\beta_{\text{diagonal}}$) as compared to all other (off-diagonal) waveguides, see Fig.~\ref{fig1}~(a).
The propagation of light across this photonic lattice is governed by the %Schr\"odinger-like 
following coupled mode equation \cite{krimer2011realization,  longhi2011photonic}
\begin{align}
i\frac{dE_{l, m}(z)}{dz}\!=\!&-\kappa \big(E_{l+1, m}+ E_{l-1, m}+ E_{l, m+1}+ E_{l, m-1}\big) \nonumber \\
&+\Delta \beta E_{l, m} \delta_{l, m} , \label{4}
\end{align}
where $E_{l, m}$ is the envelope of the electric field at the waveguide $(l, m)$, $z$ is the propagation direction, and $\kappa$ is the nearest-neighbor coupling constant. The analogy with the Hubbard problem in Eq.~\eqref{eq:single} is obtained by identifying (up to physical units) $t\!\leftrightarrow\!z$, $J\!\leftrightarrow\!\kappa$ and $U_0\!\leftrightarrow\!\Delta\beta$. 
%Although $\Delta \beta$ is negative in our experiments, it should be noted that the dynamics of the correlated particles is independent of the nature or sign of interactions~\cite{mosseri2000some, valiente2010, schneiderrosch2012}. 
In the experiment, we only excite a diagonal site and hence simulate the dynamics of two
interacting particles (either bosons or fermions with opposite spins~\cite{noba2003dynamic}) initially placed in the same potential well. 
One can, however, simulate the dynamics of two bosons or two spinless fermions placed in two different wells, by exciting two off-diagonal waveguides with symmetric ($E_{l, m}\!=\!E_{m, l}$) or anti-symmetric ($E_{l, m}\!=\!-E_{m, l}$) states respectively. Although $\Delta \beta$ is negative in our experiments, it should be noted that the dynamics of the correlated particles is independent of the nature or sign of interactions~\cite{mosseri2000some, valiente2010, schneiderrosch2012}.

%%%%%  Figure-1  (schematic diagram of photonic lattices)  %%%%%%%
%%%%%%%%%%%%% End Figure-1 %%%%%%%%%%%%%

Using this photonic simulator, we first experimentally demonstrate the suppression of standard single-particle tunneling and the emergence of tunneling of pairs, which occurs in the ``strongly-interacting" regime ($\Delta\beta \!\gg\! \kappa$). We then analyze how coherent destruction of tunneling~\cite{Longhi2012coherent} emerges for the paired particles in the presence of a simulated time-oscillating force, which is realized by periodically modulating the waveguides along the main diagonal, see Fig.~\ref{fig1}~(c). This effect includes %an on-site 
a diagonal term in Eq.~\eqref{4}, of the form  \cite{longhi2011tunneling, mukherjee2015modulation}
\begin{equation}
\frac{A n_0\omega^2 a}{\sqrt2 \lambdabar} \sin(\omega z) (l+m) E_{l,m},\label{drive:def}
\end{equation} 
where $n_0$ is the refractive index of the medium, $\lambda\!=\!2 \pi \lambdabar$ is the free-space wavelength, and $A$ [resp. $\omega$] is the amplitude [resp. frequency] of the sinusoidal modulation. Note that this modulation corresponds to adding a time-dependent driving term 
\begin{equation}
\hat W(t)\!=\!K\sin(\omega t) \sum_s s \hat{n}_s \label{2},
\end{equation}
in the original Hubbard Hamiltonian [Eq.~\eqref{Hubbard}], with the simple identification $K\!\leftrightarrow\!  A n_0\omega^2 a/(\sqrt2 \lambdabar)$.
%%%%%%%  Time evolution of two interacting particles %%%%%%%

%%%%%%  Figure-2  Intensity distribution photonic lattices  %%%%%%%%
\begin{figure}[t]
\includegraphics[width=8.6 cm]{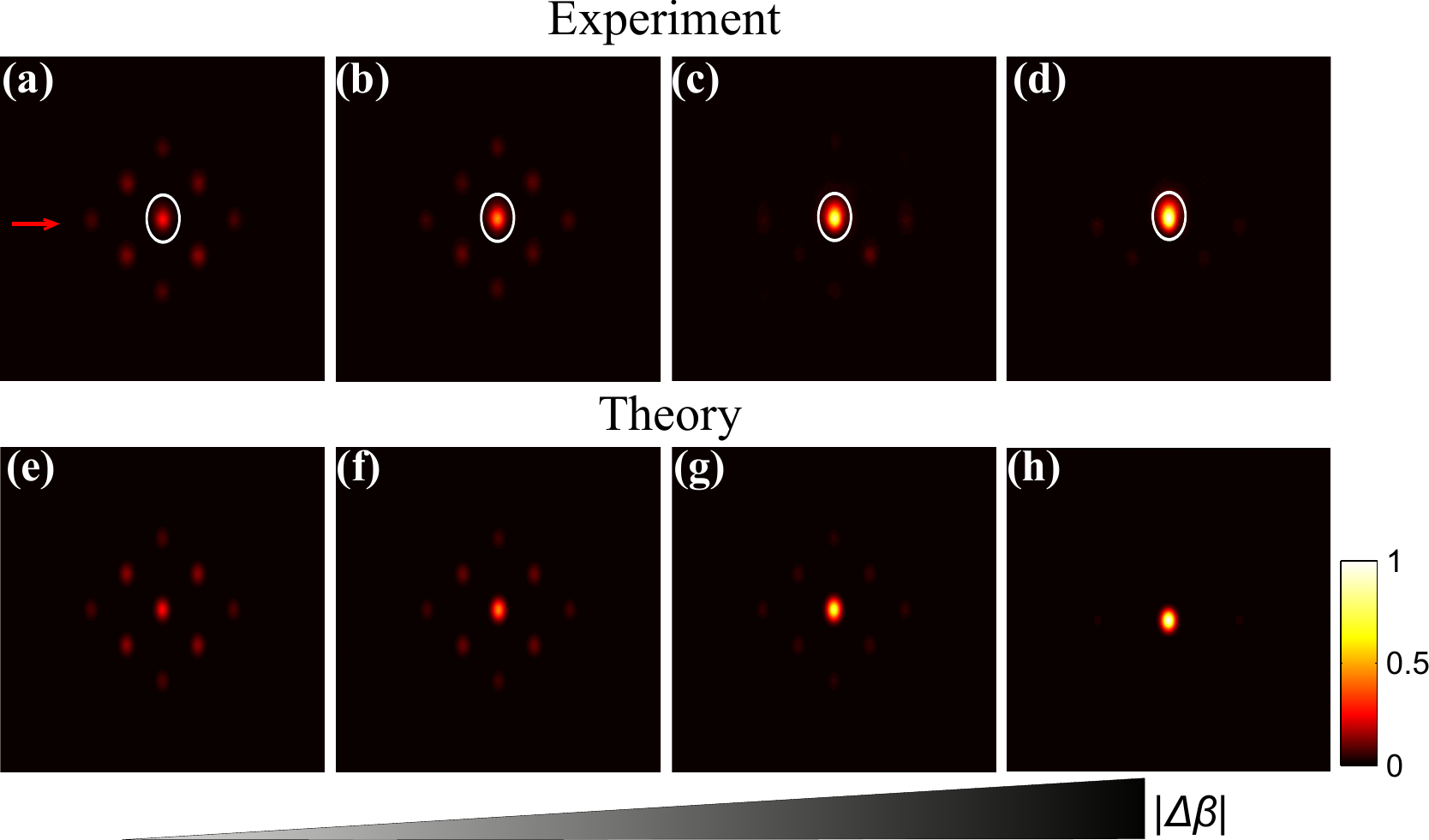}
\caption{Experimentally observed output intensity distributions when light was launched into the $l, m=4, 4$ site (circled waveguide) of the 15-mm-long straight photonic square lattices ($A\!=\!0$). Here the coupling constant is $\kappa\!=\!0.04$~mm$^{-1}$. The analogous interaction $\Delta \beta$ is (a) $0.0$, (b) $-0.12$~mm$^{-1}$, (c) $-0.21$~mm$^{-1}$ and (d) $-0.32$~mm$^{-1}$. The corresponding simulated intensity distributions are shown in (e)-(h). Each image is normalized such that the total output power is 1. See also Fig.~\ref{fig2S}. %[S2] %\ref{fig2S} in Ref.~\cite{suppmat}.
The red arrow indicates the main diagonal ($l\!=\!m$).
}
\label{fig3}
\end{figure}
%%%%%%% END Fig-2  %%%%%%%%%%%%%%%%%%%%%%%%%%%

{\it Time evolution of two interacting particles.} 
In order to observe the effect of the on-site interaction,
we fabricated ten  straight ($A\!=\!0$) square lattices of 15~mm propagation length ($z_{max}$) with ten different values for the shifted propagation constant, $\Delta\beta$.  
The precise tuning of $\Delta\beta$ was realized, without affecting $\kappa$ and waveguide losses, by changing the translation speed ($v$) of fabrication; within the range of our fabrication parameters, the shift in propagation constant varies almost linearly with translation speed~[Fig.~\ref{fig1S}]; %\cite{suppmat}; 
see also~\cite{corrielli2013fractional}.
The lattice constant is $a\!=\!16.35$~$\mu m$, the nearest-neighbor coupling constant is $\kappa\!=\!0.04$~mm$^{-1}$ and the next-nearest-neighbor coupling was insignificant for the maximum observable propagation distance (for all lattices).
The off-diagonal waveguides have, in all cases, the same propagation constant, since they were fabricated with the same translation speed ($v_{od}\!=\!9$~mm/s). The diagonal waveguides in each lattice were fabricated with a different translation speed, $v_d\!=\!v_{od}-\Delta v$, with $\Delta v\!=\!0$ to 4.5~mm/s in steps of 0.5~mm/s, resulting in a different propagation constant ($\Delta\beta$) along the diagonal; this gives rise to the analogous on-site ``interaction" term [Eq.~\eqref{4}]. Fig.~\ref{fig3}~(a) shows the observed output intensity distribution for $\Delta v\!=\!0$ (or $\Delta \beta\!=\!0$) and for the input condition: $E_{4, 4}(0)\!=\!1$. 
Applying the same input condition, we observed that the output intensity distribution becomes increasingly localized as $\Delta v$ was increased [Fig.~\ref{fig3}~(b)-(d)]; this is in excellent agreement with the %corresponding 
theoretical intensity distributions [see Fig.~\ref{fig3}~(e)-(h)], which were obtained by solving the coupled-mode equations (Eq.~\ref{4}). Theoretically, these results  can be understood in terms of the effective mass of the two-body bound state associated with the Hubbard Hamiltonian Eq.~\eqref{Hubbard}, i.e.~$M^*\!\approx\!-\hbar^2U_0/(4J^2a^2)$, whose absolute value increases as a function of the interaction strength in the strongly-interacting regime~\cite{valientepetrosyan2008}. Note that the values of $\kappa$ and $z_{max}$ were chosen such that the lattice sites at the edges are never excited; see Fig.~\ref{fig2S}. %[S2] %\ref{fig2S} in \cite{suppmat}.

%%%%%%%%%%%%  Figure -3 pair tunneling %%%%%%%%%%%%%%
\begin{figure}[]
\includegraphics[width=8.6 cm]{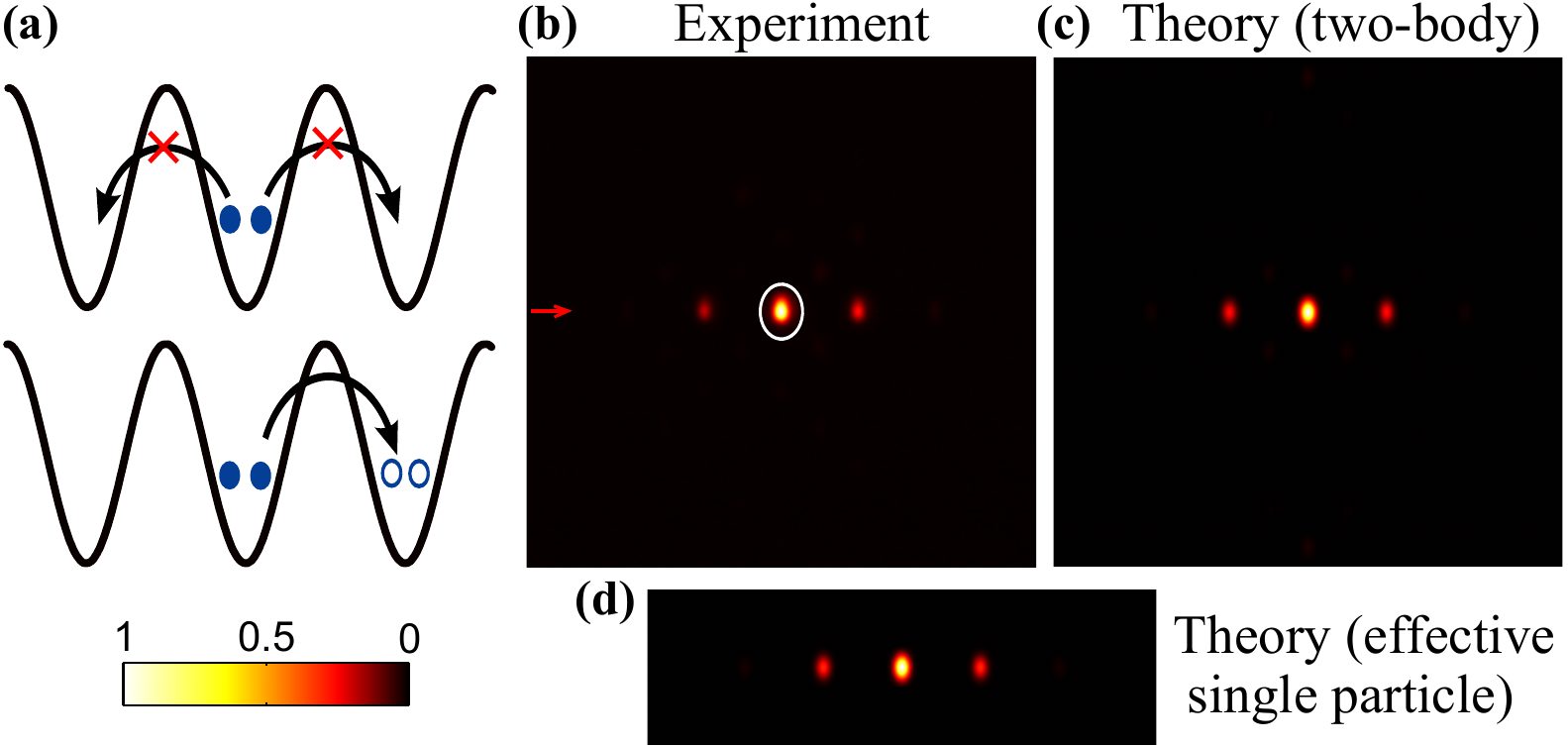}
\caption{(a) Schematic diagram of pair hopping in the strong-interaction regime [$U_0\!\gg\! J$]. 
(b) Experimental observation of pair hopping in the 70-mm-long straight photonic square lattices for $\kappa\!=\!0.027$~mm$^{-1}$, $\Delta \beta\!=\!-0.21$~mm$^{-1}$. %($v_d=6$~mm/s). 
The corresponding two-body simulation and the effective single-particle simulation with pair coupling $\kappa_2\!=\!-2\kappa^2/\Delta\beta$, are shown in (c) and (d), respectively. Each image is normalized such that the total output power is 1. The white circle indicates the waveguide that was excited at the input.}
\label{fig4}
\end{figure}
%%%%%%%%%%%%  END Figure -3  %%%%%%%%%%%%%%%%%

In the strongly-interacting regime of the Hubbard model, standard tunneling for individual particles is replaced by two-body tunneling processes. Such a pair tunneling cannot be identified in Fig.~\ref{fig3}, due to the very weak pair tunneling parameter $\kappa_2\!\ll\! \kappa$ and the small propagation length (15~mm) of the lattices. Hence, in order to reveal pair tunneling, a longer photonic lattice (70-mm-long) was fabricated; note that a larger lattice constant (18~$\mu m$) was also chosen in order to inhibit next-nearest-neighbor hopping in this longer lattice. Here, the other lattice parameters are given by $\kappa\!=\!0.027$~mm$^{-1}$ and $\Delta \beta\!=\!-0.21$~mm$^{-1}$.
When the (4, 4) waveguide was excited at the input of this lattice, we observe significant amount of light in the diagonal waveguides [Fig.~\ref{fig4}~(b)]. This is a direct demonstration of pair tunneling, and hence, the formation of a bound state in the strongly-interacting regime. We see good agreement between experimental results, the full two-body simulation [i.e. the solution of Eq.~(\ref{4})] and an effective single-particle simulation; the latter corresponds to the dynamics of a paired state hopping in a one-dimensional lattice, with the modified (pair tunneling) constant $\kappa_{2}\!=\!-2\kappa^2/\Delta \beta$~\cite{valientepetrosyan2008, petrosyan2007quantum}.

%%%%%%%%%%%%%  Coherent destruction of tunneling   %%%%%%%%%
{\it Coherent Destruction of Tunneling (CDT).} 
The dynamics of the two interacting particles in the presence of a sinusoidal driving force is determined by  $\kappa$, $\Delta\beta$, $\omega$ and $K$ [Eq.~\eqref{eq:single}-\eqref{2}], and the transport along the array can be suppressed (CDT) under the following conditions.
In the high frequency regime ($\omega\!\gg\!\Delta\beta$), the external driving renormalizes the hopping amplitude as for the non-interacting case and causes approximate CDT when the zero-th order Bessel function [$\mathcal{J}_0(K/\omega)$] is zero. For $\omega\!\ll\!\Delta\beta$ (low frequency regime) CDT can not be observed simultaneously for paired and unpaired particles as the effective force on the paired particle is twice the force acting on an unpaired particle. In this context the role of $K/\Delta \beta$ was elaborated in Ref.~\cite{creffield2004localization}; see also \cite{noba2003dynamic}.
Recently, it was reported in Ref.~\cite{Longhi2012coherent} that CDT can be simultaneously realized for both paired and unpaired particles under appropriate driving if $\omega\!\sim\!\Delta\beta$ and $\omega \gg \kappa$, up to the long time scale ($\sim \omega/\kappa^2$).

\begin{figure}[]
\includegraphics[width=8.6 cm]{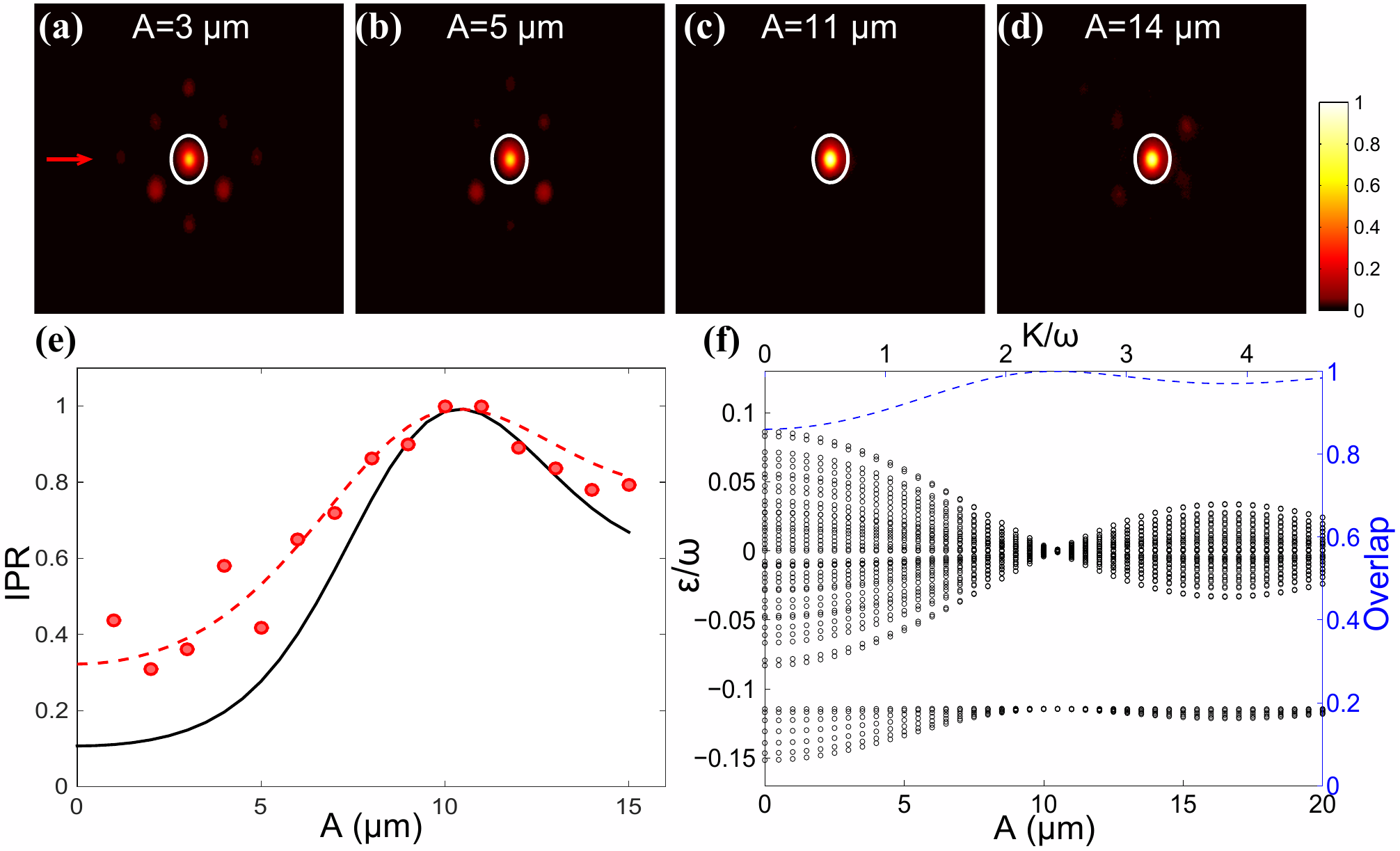}
\caption{(a)-(d) The intensity distributions at the output of modulated square lattices (15-mm-long) for the input condition: $E_{4, 4}(0)\!=\!1$.
The amplitude of modulation ($A$) is indicated for each image. A different color-scale was chosen to highlight the low intensities.
(e) The variation of IPR as a function of $A$.  
The red circles show the measured IPR. The dotted red line indicates the theoretically obtained variation of IPR. Here $\Delta \beta\!=\!-0.18$ mm$^{-1}$, $\kappa\!=\!0.04$  mm$^{-1}$, and $\omega\!=\!1.57$ ~mm$^{-1}$. The solid black curve represents the variation of IPR for $\Delta\beta=\!0$. 
(f) Floquet quasienergies (Black) for a sinusoidally driven one-dimensional lattice with two interacting particles.  
The wider miniband corresponds to the unpaired state and the narrower miniband is for the paired states. 
The overlap of the input state, $E_{4, 4}\!=\!1$ and the eigenstates of the narrower miniband as a function of $A$ is shown by the dotted blue line.
}
\label{fig5}
\end{figure}

Here we focus on CDT in the high frequency regime ($\omega\! \gg \! \Delta\beta, \kappa$). In the experiment, fifteen square lattices (15-mm-long) were fabricated with sinusoidally modulated waveguides [Fig.~\ref{fig1} (c)]. The amplitude ($A$) of modulation was varied from 1 to 15~$\mu m$ in steps of 1~$\mu m$. 
The on-site interaction $\Delta \beta\!=\!-0.18$~mm$^{-1}$, driving frequency $\omega\!=1.57$~mm$^{-1}$ and the hopping amplitude, $\kappa\!=\!0.04$~mm$^{-1}$ in all cases.
Fig.~\ref{fig5}~(a)-(d) show the output intensity distributions when light is launched at the (4, 4) site of each lattice. As $A$ is increased the output intensity distribution is observed to be increasingly localized and becomes almost fully localized near $A=11$~$\mu m$.  As $A$ was further increased, the tunneling is restored. 
To quantify the localization, the inverse participation ratio (IPR) was calculated from the
measured output intensity distributions. The IPR is a measure of localization and is defined as~\cite{mukherjee2015modulation}: $\text{IPR}=\sum I_{l,m}^2/\big(\sum I_{l,m}\big)^2$, where $I_{l,m}$ is the light intensity at the ($l$, $m$) waveguide. The variation of IPR as a function of $z$ is shown in Fig.~\ref{fig3S}. %[S3] %\ref{fig3S} in~\cite{suppmat}.
The measured IPR (at $z\!=\!15$~mm) as a function of $A$ is shown in Fig.~\ref{fig5}~(e) by the red circles, in good agreement with the theoretical result (dotted red line). To highlight the difference between the observed IPR for two interacting particles and that for two non-interacting particles, the theoretical IPR with no interactions ($\Delta\beta=0$) is shown by the solid black line for which CDT occurs when $\mathcal{J}_0(K/w)\!=\!0$ (i.e.~$A\!\approx\!11$~$\mu$m).
Note that site-dependent loss is not important for our lattices~(see APPENDIX). %\cite{suppmat}.
%%%%%%%%%%%%  Figure -4 Modulated lattices %%%%%%%%%%%%%
%%%%%%%%% End Figure -4 Modulated lattices %%%%%%%%%%%%%%%

The observed variation of IPR with the amplitude of modulation can be explained from the Floquet quasienergy spectrum \cite{holthaus1993quantum, creffield2004localization}.
For the modulated square lattices, the Hamiltonian is a periodic function of $z$, i.e. $\hat H(z)\!=\!\hat H(z+z_0)$. In this situation, using Floquet theory, the solution of the Schr\"odinger equation can be written as $\ket {\Phi(z)}\!=\!e^{-i\varepsilon z} \ket {\phi(z)}$ with $\ket {\phi(z)}\!=\!\ket {\phi(z+z_0)}$, where $\ket {\phi(z)}$ is a Floquet state and $\varepsilon$ is the quasienergy. To obtain the quasienergies we diagonalize the evolution operator defined as:
\begin{eqnarray}
U= \mathcal{T} \exp\Big[-i \int_0^{z_0} H(z') dz' \Big]   \label{7}
\end{eqnarray}
where $\mathcal{T}$ indicates the time ordering. The normalized quasienergy ($\varepsilon/\omega$) is plotted in Fig.~\ref{fig5}~(f) as a function of $A$ for fixed values of $\Delta\beta$ and $\kappa$. 
As can be seen in Fig.~\ref{fig5}~(f), the quasienergies for two interacting particles in a  periodically-driven one dimensional lattice form two minibands.
The wider miniband corresponds to the states in which the two particles are in separate wells (unpaired state) and the narrower miniband corresponds to the paired states. The minibands are separated by a gap $\sim \Delta \beta$, corresponding to the (anti-) binding energy. Both minibands (pseudo) collapse near $A\!=\!11$ $\mu$m which causes (approximate) CDT. The calculation of the overlap of the input state, $E_{4, 4}\!=\!1$ and the eigenstates of the narrower miniband as a function of $A$ is shown by the dotted blue line in  Fig.~\ref{fig5}~(f). The overlap is more than 80\% irrespective of the value of $A$. Importantly, since the narrower miniband is less dispersive compared to the  
Floquet band of two non-interacting particles (not shown in the figure), the localization (IPR) of the interacting system is stronger than in the case of a non-interacting system (i. e. the dotted red line in Fig. \ref{fig5}~(e) lies above the solid black line).

{\it Conclusions.} 
We have experimentally implemented the photonic realization of two interacting particles in a one-dimensional tight-binding lattice with only nearest-neighbor tunneling. The suppression of independent tunneling of individual particles and the evidence of second order pair tunneling was observed in the strong interaction regime. We then showed the effect of an engineered sinusoidal force on the paired state. Coherent destruction of tunneling was observed as the amplitude of the force is varied. Our experiment paves the way for simulating few-body physics in low-dimensional lattices with a clean nearest-neighbor-only hopping, for both static and periodically-driven (or Floquet) Hamiltonians. Successful implementation of similar photonic setups will enable us to experimentally simulate other intriguing problems such as {\it N} particles in a double-well potential~\cite{longhi2011optical}, the dynamics of the correlated particles in a random potential~\cite{krimer2011two, albrecht2012induced}, dissipation-induced correlation~\cite{rai2015photonic} and two-body Su-Schrieffer-Heeger model~\cite{di2016two}.

\vspace{0mm}

\newcommand{\beginsupplement}{%
        \setcounter{equation}{0}
        \renewcommand{\theequation}{A\arabic{equation}}%
        \setcounter{figure}{0}
        \renewcommand{\thefigure}{A\arabic{figure}}%
     }
\beginsupplement

\section*{APPENDIX}
%%%%%%%%%%%%%%%%%%%%%%%%%%%%
In this appendix, we briefly discuss the fabrication parameters, measurement of coupling constant and analogous interaction (or shift in propagation constant) for completeness. We then present the measured values of waveguide losses (propagation loss+bend loss) and finally show numerical simulations related to Fig.~\ref{fig3}~(a) and Fig.~\ref{fig5}~(a)-(d).
We show that within the range of our fabrication parameters $\Delta\beta$ varies linearly with translation speed (see also Ref.~\cite{corrielli2013fractional, szameit2009inhibition}) without affecting $\kappa$ and waveguide losses (see also Ref.~\cite{heinrich2014supersymmetric}).

{\it Fabrication and characterizations.} 
The coupled optical waveguide arrays were fabricated using the femtosecond laser inscription technique~\cite{davis1996writing}. 
The substrate material (Corning Eagle$^{2000}$) was mounted on $x$-$y$-$z$ translation stages (Aerotech: ABL1000) and each waveguide was fabricated by translating the substrate once through the focus of a 500 kHz train of circularly polarized sub-picosecond ($\sim$350 fs) laser pulses, generated by a Menlo BlueCut fiber laser system. The waveguide refractive index profile was controlled using the  slit-beam shaping method~\cite{ams2005slit}. 
The white-light-transmission micrograph of the facet of a finite square lattice is shown in Fig.~\ref{fig1}~(d). The lattice contains 49 waveguides, each of which supports only a fundamental mode at 740 nm wavelength. 
The lattice axes were rotated by 45$^o$ with respect to the vertical direction (V) to achieve $\kappa_x \approx \kappa_y$, where $\kappa_x$ and $\kappa_y$ are the coupling strengths along the two axes of the lattice.

\begin{figure}[]
\includegraphics[width=8.6 cm]{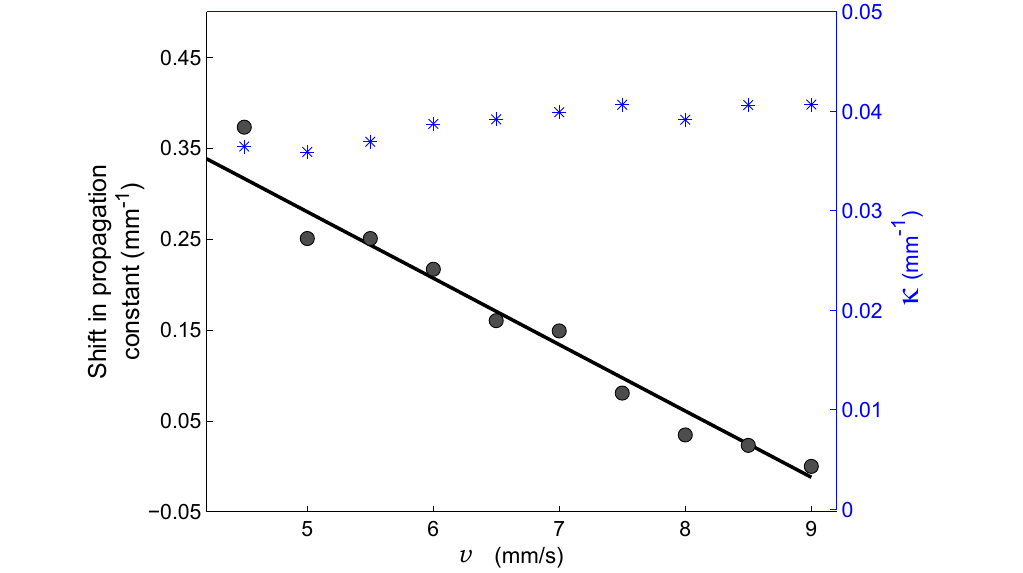}
\caption{(Black) Variation of shift in propagation constant [$\beta(v)-\beta(v=9~{\text{mm/s}})$] as a function of translation speed ($v$). The black circles are the measured values and the solid line is the linear fit. (Blue) The variation of hopping amplitude ($\kappa$) with translation speed can be ignored as it is comparable to the random variation of $\kappa$ (i.e. the off-diagonal disorder).  
}
\label{fig1S}
\end{figure}

The parameters that describe a photonic lattice are the coupling constants between the lattice sites and the propagation constant of each waveguide. 
To estimate the variation of nearest-neighbor coupling constant ($\kappa$) with translation speed, ten symmetric evanescently coupled two-waveguide couplers (both waveguides with identical fabrication parameters) were fabricated with translation speeds 9 to 4.5~mm/s in steps of 0.5~mm/s. As can be seen from Fig.~\ref{fig1S}, the variation of $\kappa$ with translation speed can be ignored as it is comparable to the random variation of $\kappa$ (i.e. the off-diagonal disorder). 
To measure next-nearest neighbor coupling ($\kappa_n$), two-waveguide evanescent couplers were fabricated with waveguide-to-waveguide separation $D=\sqrt{2} a$, $a$ is the lattice constant of the square lattices, and it was found that $\kappa_n \approx 0$ for the maximum observable propagation distance. 
To estimate the variation of the waveguide refractive index ($n_{wg}$) with translation speed, we fabricated ten one dimensional diffraction gratings with  15~$\mu m$ grating period, using translation speeds of 9 to 4.5 mm/s. 
The physical depth of each grating was measured (using an optical microscope) to be $d= 5\pm 0.5$ $\mu m$, and it is reasonable to assume that for our inscription parameters, the variation of writing speed only changes the magnitude of the refractive index contrast, not its spatial extent. By measuring the absolute diffraction efficiency of these gratings at the first order ($\eta_1$), and assuming that the diffraction gratings are sinusoidal phase gratings, $n_{wg}$ is calculated from~\cite{martinez2007fabrication}, as:
\begin{equation}
\eta_1=\mathcal{J}_1^2\Big(\frac{2\pi(n_{wg}-n_{0})}{\lambda}d\Big) \label{5}
\end{equation}
where $n_{0}$ is the refractive index of the substrate, and $\mathcal{J}$ is the Bessel function of the first kind. The quantity of our interest is the shift in propagation constant (i.e. $\!\frac{2\pi}{\lambda} \Delta n_{\text{eff}}$) as the translation speed is changed. 
Assuming that the difference in the effective indices of the modes ($\Delta n_{\text{eff}}$) is very close to the difference in waveguide refractive indices ($\Delta n_{wg}$) for two waveguides fabricated with two different translation speeds, we plot the shift in propagation constant [$\beta(v)-\beta(v=9~{\text{mm/s}})$] as a function of translation speed; see Fig.~\ref{fig1S}. 
The black circles in Fig.~\ref{fig1S} are the measured values and the solid line is the linear fit.

%%%%%%%%%%%%  Figure -1-supp  %%%%%%
%%%%%%%%%%%%%%% end Fig-1-supp %%%%%%%%%%%%

%%%%%%%%%%%%  Figure -2-supp  %%%%%%
\begin{figure}[]
\includegraphics[width=8.5 cm]{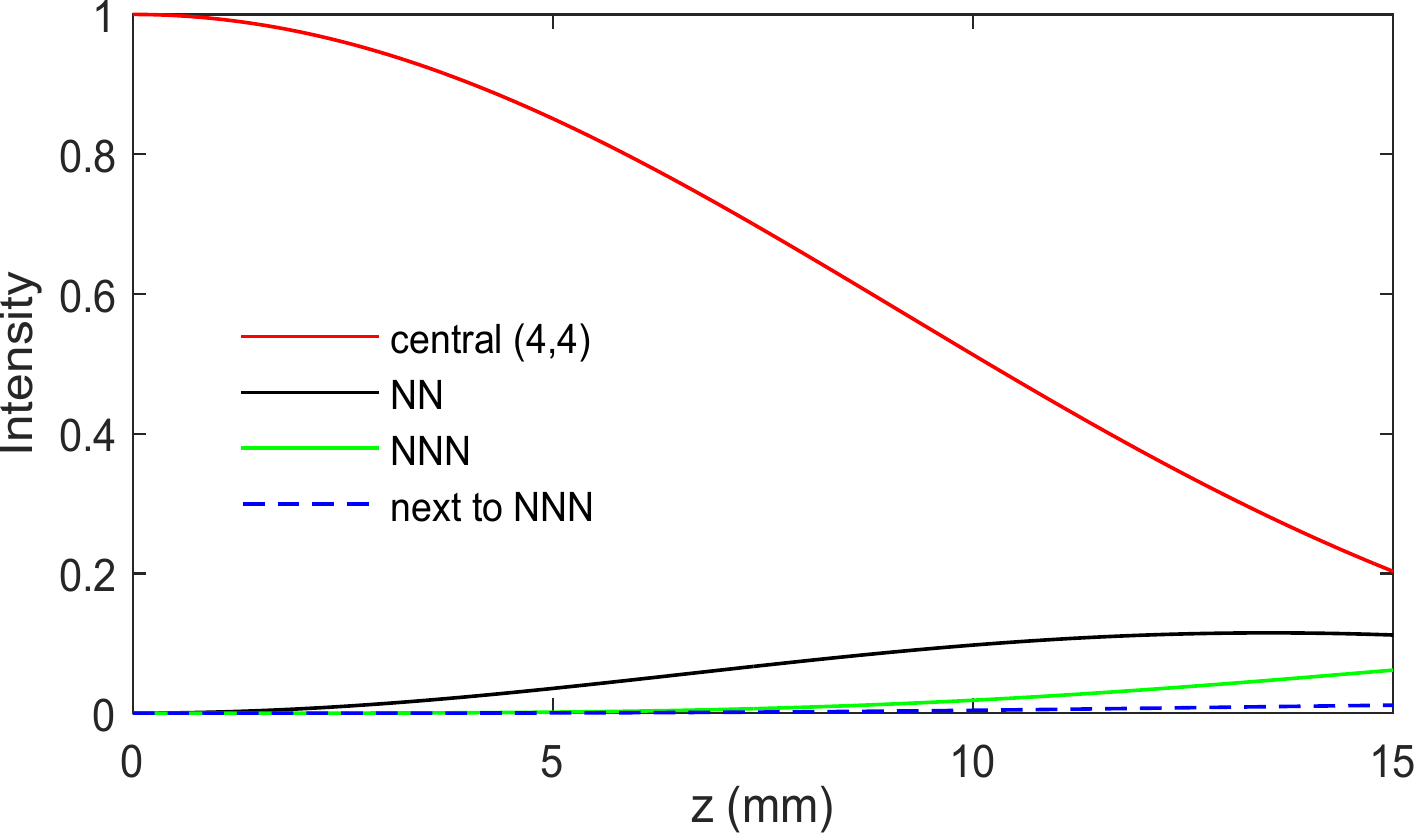}
\caption{Variation of light intensity along the propagation direction, $z$, for a photonic square lattice. Here $\Delta \beta\!=\!0$, $A\!=\!0$ and $\kappa\!=\!0.04$~mm$^{-1}$. The red curve is for the central waveguide that was excited at the input ($z\!=\!0$). The black, green and the dotted blue curves indicate intensities at the nearest neighbor (NN), next-nearest neighbor (NNN) and next to next-nearest neighbor waveguides. Fig.~\ref{fig3}~(a) shows the measured output intensity distribution at $z\!=\!15$~mm.}
\label{fig2S}
\end{figure}
%%%%%%%%%%%%%%% end Fig-2-supp %%%%%%%%%%%%
%%%%%%%%%%%%  Figure -3-supp  %%%%%%
\begin{figure}[h!]
\includegraphics[width=8.5 cm]{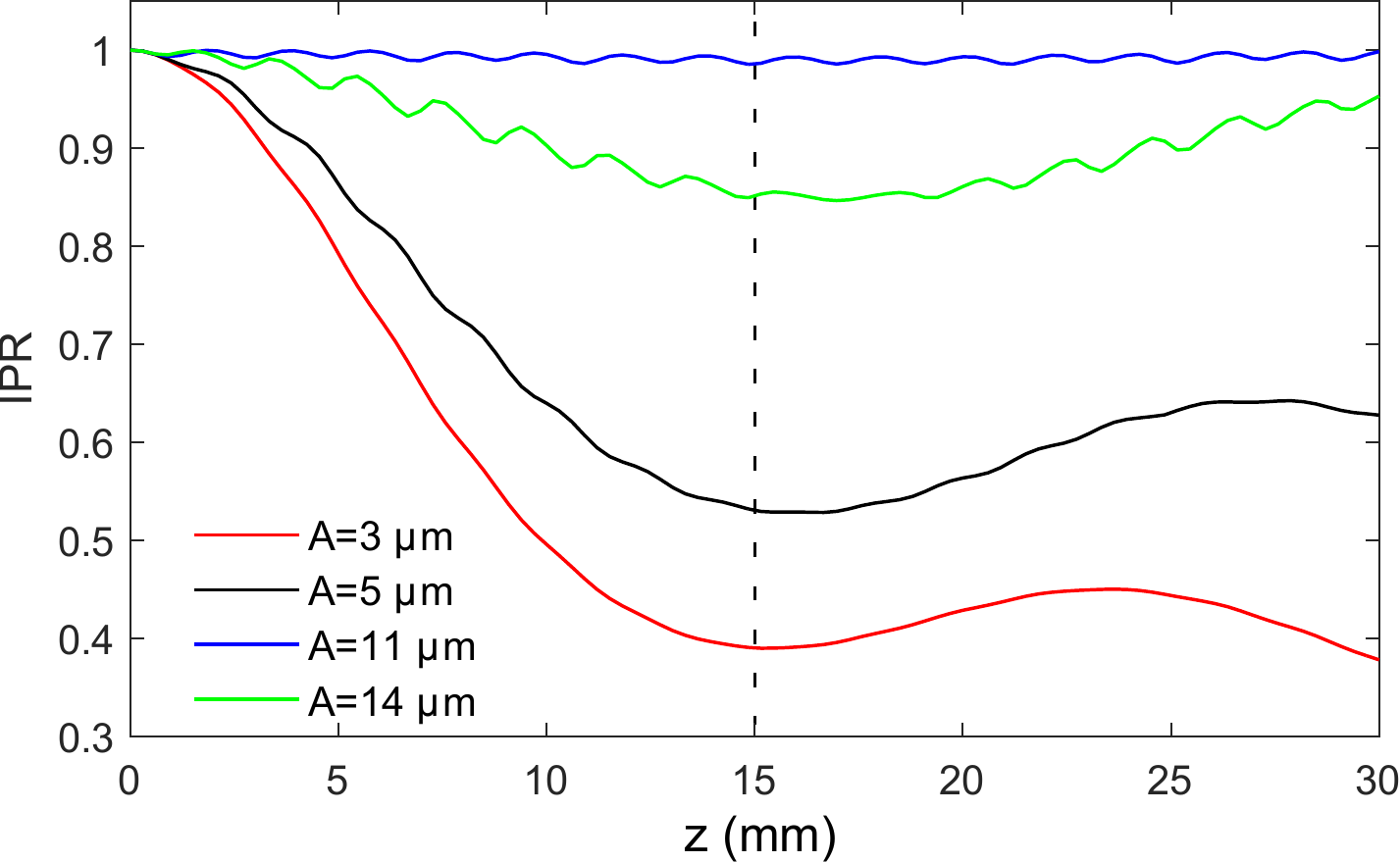}
\caption{Variation of inverse participation ratio (IPR) as a function of propagation distance, $z$, for the modulated photonic lattices presented in Fig.~\ref{fig5}~(a)-(d). The dotted line indicates the $z$ value where the output intensities were experimentally measured.}
\label{fig3S}
\end{figure}
%%%%%%%%%%%%%%% end Fig-3-supp %%%%%%%%%%%%

{\it Loss measurement.} 
To measure the variation of propagation loss with translation speed, ten isolated straight waveguides (translation speeds 9 to 4.5~mm/s in steps of 0.5~mm/s) were fabricated in a 15-mm-long substrate. We estimated propagation loss by subtracting the coupling losses from the measured insertion loss for each waveguide. The propagation loss of the waveguide fabricated with 9~mm/s was found to be 0.92~dB/cm. The variation of propagation loss with translation speed was insignificant, with a maximum fluctuation of $\approx$~0.04~dB/cm. 
To estimate bend loss, two sets of sinusoidally modulated waveguides were fabricated with translation speeds of 9 and 6.5~mm/s. For each set, 15 waveguides were fabricated with 1 $\mu m \le A \le$ 15~$\mu m$ in steps of 1~$\mu m$.
Now the quantity of interest is the difference in total loss (propagation loss+ bend loss) for two waveguides fabricated with equal amplitude of modulation and two different translation speeds, which was measured to be $<0.15$~dB for 15-mm-long waveguides. Therefore the site-dependent loss can be ignored.

{\it Numerical simulations.} 
In Fig.~\ref{fig2S}, we have shown the numerically calculated evolution of intensity distribution as a function of $z$ for a $7\!\times \!7$ photonic square lattice with $\Delta\beta\!=\!0$ and $A\!=\!0$, for which the effective tunneling is maximal, see also Fig.~\ref{fig3}~(a). 
It should be mentioned that for all the experiments presented in the main text, the waveguides at the edges are not excited.
We then show, Fig.~\ref{fig3S}, how the inverse participation ratio (IPR) varies as a function of $z$ for four different values of $A$.
Note that the evolution of light intensity along the propagation direction can be experimentally measured by detecting the fluorescent emission if the waveguide arrays are fabricated inside fused silica instead of Corning Eagle$^{2000}$; see Ref.~\cite{corrielli2013fractional}.

%%%%%%%%%% BH-bib %%%%
%\bibliography{BH-bib}

\bigskip 

{\it Acknowledgements.} 
S.~M. thanks Heriot-Watt University for a James Watt Ph.D Scholarship. N.~G. is financed by the FRS-FNRS Belgium and by the BSPO under the PAI project P7/18 DYGEST. R.~R.~T. gratefully acknowledges funding from the UK Science and Technology Facilities Council (STFC) in the form of an STFC Advanced Fellowship (ST/H005595/1).   % input acknowledgement

%%%%%%%%%% BH-bib %%%%
%\bibliography{BH-pra-bib}

\end{document}